\documentclass[sigconf]{acmart}

\setcopyright{none}
\renewcommand\footnotetextcopyrightpermission[1]{}

\usepackage{xspace}    \usepackage{array}     \usepackage{multirow}  \usepackage{enumitem}  \usepackage{xcolor}    \usepackage{placeins}  
\newcommand{\sysname}{\textsc{DBA-Bench}\xspace}

\newcommand{\baseline}[1]{\textsf{#1}}

\newcommand{\lineit}[1]{\noindent\underline{\textit{#1}}}

\newcommand{\topic}[1]{\lineit{#1}}

\AtBeginDocument{}

\begin{document}

\title{\sysname: A Production-Fidelity Benchmark for
  LLM-Based Database Operations Agents}

\author{Junming~Chen}
\affiliation{\institution{University of Electronic Science and Technology of China}
  \city{Chengdu}
  \country{China}
}
\email{junmingchen@std.uestc.edu.cn}

\author{Junyang~Jiang}
\affiliation{\institution{University of Electronic Science and Technology of China}
  \city{Chengdu}
  \country{China}
}
\email{junyangjiang@std.uestc.edu.cn}

\author{Xu~Chen}
\affiliation{\institution{University of Electronic Science and Technology of China}
  \city{Chengdu}
  \country{China}
}
\email{xuchen@std.uestc.edu.cn}

\author{Zibo~Liang}
\affiliation{\institution{University of Electronic Science and Technology of China}
  \city{Chengdu}
  \country{China}
}
\email{zbliang@std.uestc.edu.cn}

\author{Kai~Zheng}
\affiliation{\institution{University of Electronic Science and Technology of China}
  \city{Chengdu}
  \country{China}
}
\email{zhengkai@uestc.edu.cn}

\renewcommand{\shortauthors}{Chen et al.}

\begin{abstract}

LLM-based database agents show promise, but differences in task scope,
testbeds, and success criteria hinder direct comparison. A shared benchmark
must also preserve the operational conditions that determine whether
diagnosis and remediation succeed in practice. We present \sysname, guided by
\emph{operational realism}, \emph{outcome-first evaluation}, and
\emph{controlled scenario reproducibility}. It evaluates agents in live
PostgreSQL and distributed TiDB deployments with active workloads, stateful
behavior, and multi-source observations. Success requires satisfying
scenario-specific operational objectives without violating safety
constraints. The primary evaluation covers 231 bounded scenarios across five
operational domains, while 10 persistent scenarios extend evaluation to
continued operation across interacting events.
We evaluate eight automated baselines, including four frontier foundation
models, two open-weight models, and two GPT-5.5-backed database agents.
Our evaluation exposes gaps between diagnosis, effective remediation, and
safe execution across both database systems. Agents can identify root causes
yet fail to restore operational objectives. Even when objectives are restored,
the executed actions can violate safety constraints. Persistent scenarios
expose a gap between clearing individual blockers and sustaining business
service: agents leave competing workloads uncoordinated and previously
completed repairs unchecked after subsequent changes.

\end{abstract}



\maketitle
\pagestyle{plain}

\section{Introduction}
\label{sec:intro}

Database operations encompass routine maintenance, incident response, 
and planned changes needed to maintain the availability, performance, 
and data integrity of database services. These tasks are carried out 
under ongoing workloads and service constraints, requiring database 
administrators (DBAs) to account for the operational consequences of 
their interventions. Recent LLM-based systems have begun to assist with 
such tasks. For example, D-Bot uses a multi-expert diagnosis framework 
based on tree search~\cite{Zhou2024DBot}, DBAIOps guides diagnostic paths 
with a knowledge graph~\cite{Zhou2026DBAIOps}, and DBAgent uses reinforcement
learning to improve diagnostic reasoning~\cite{Chen2026DBAgent}.

Across these studies, however, evaluations differ in task scope, testbed, 
and success criteria~\cite{Zhou2024DBot,Zhou2026DBAIOps,Chen2026DBAgent}, 
making their reported results difficult to compare directly.
A shared benchmark is needed to place these systems on a common evaluation
basis. To make that comparison informative about how agents perform in
practice, the benchmark must also preserve the operational conditions that
shape diagnosis and remediation in production. We consider four such aspects:

\textbf{(A1) Live-environment fidelity.}
In production, database operations take place in a running, writable system.
Workloads continue during diagnosis and remediation. Interventions can change
locks, optimizer state, availability, and the evidence available for subsequent
decisions. Existing database-agent evaluations often emphasize diagnosis or
recommendation quality. Proposed fixes are assessed through reports or estimated
benefits~\cite{Zhou2024DBot,Zhou2026DBAIOps}. Such assessments do not establish
whether the fixes work under an active workload.

\textbf{(A2) Observation-space scale and complexity.}
A production incident is not isolated from the rest of the system.
DBAs face large volumes of metrics, business logs, query plans, and lock and
session records. Causal evidence is mixed with signals from concurrent
workloads and background tasks. Compact task environments that omit this
competing activity simplify the investigation. They leave causal localization
amid operational noise untested.

\textbf{(A3) Open solution space.}
Real-world remediation rarely has a single correct answer. Competing fixes
often satisfy the same technical objective while imposing different
availability and maintenance costs. For example, an agent can address a slow
query by adding an index. In PostgreSQL, a standard index build blocks writes
to the table until it completes. A concurrent build allows writes to continue
but generally takes longer~\cite{PostgreSQL2026Docs}.
Scoring against a fixed solution can reject valid alternatives. Judging only
the final performance can overlook service disruption during the intervention.

\textbf{(A4) Scenario complexity and coverage.}
Production database operations span tasks with different objectives and
causal structures. Query tuning, system failure, health check, business change,
and resource governance demand different operational decisions. Within these
tasks, faults may interact across database mechanisms:
stale statistics can degrade a query plan, amplify lock contention, and exhaust
a connection pool. Success on isolated faults within a narrow task domain
provides limited evidence of broader operational capability.

We present \sysname, a shared benchmark that turns these requirements into
three design principles.

\textbf{(P1) Operational Realism.}
\sysname instantiates operational tasks in running PostgreSQL and distributed
TiDB deployments. OLTP, OLAP, or mixed workloads remain active as agents
execute interventions that change database state. The tool interface exposes
database telemetry and operational records, retaining competing signals from
concurrent workloads and background processes. The scenario library spans
five operational domains and includes faults that interact across database
mechanisms.

\textbf{(P2) Outcome-First Evaluation.}
Scenario-specific verifiers check task completion end to end using the
resulting database state and required task outputs. Different remediation
paths receive credit when they achieve the operational objective. Separate
safety checks assess the action trace and resulting state against the
scenario's operational constraints.

\textbf{(P3) Controlled Scenario Reproducibility.}
For individual tasks, \sysname restores the prepared environment and rechecks
the intended fault conditions and visible symptoms before each run. Agents
therefore face equivalent fault conditions, while workload activity and their
interventions continue to shape the database state.

Our primary comparison uses \emph{bounded scenarios}, which assess whether an
agent can complete one operational task end to end, starting from a restored
environment with an active workload. The corpus contains 107 PostgreSQL and 124
TiDB scenarios across five operational domains: query tuning, system failure,
health check, business change, and resource governance. Each scenario
carries a public Easy or Hard label based on reference-path diagnostic depth
and environmental complexity.

As a complement to this per-task setting, we introduce \emph{persistent
scenarios} to evaluate agents during continued database operation. The database
and workload run continuously for 1--2~hours while operational events are
injected over time. Later events may interact with state left by earlier
interventions or with unresolved earlier events, testing whether the agent can
maintain service stability throughout the run. The persistent corpus contains
5 PostgreSQL and 5 TiDB scenarios.

We evaluate eight automated baselines: four frontier foundation models, two
open-weight models, and two GPT-5.5-backed database agents.
DBAIOps achieves the highest observed Safe Pass on PostgreSQL at 27.1\%,
while Claude Opus 4.8 leads on TiDB at 21.0\%. The pass decomposition
reveals distinct gaps within end-to-end operation. Among PostgreSQL runs that
pass Diagnosis, 72.2\% still fail Outcome Pass, indicating that correct causal
localization often does not translate into effective remediation. Among runs
that restore the required outcome, 17.3\% fail Safe Pass because the action
trace or resulting state violates an operational constraint. In the independent
TiDB diagnosis cohort, 71.6\% of diagnosis-passing runs fail Outcome Pass.
In the TiDB outcome evaluation, 16.1\% of outcome-restoring runs fail Safe Pass.
These conditional rates use run-level verdicts; diagnosis comparisons are
limited by unequal coverage of resolved, applicable semantic judgments.
Across four models and ten persistent scenarios, Service Compliance passes
in 15/40 runs, complete Outcome Pass in 3/40, and Safe Pass in 2/40.
These small cohorts characterize performance on the tested event sequences.

In summary, this paper makes three contributions:
\begin{itemize}[leftmargin=*,nosep]
\item \textbf{A benchmark for comparable, operationally grounded evaluation.}
We develop \sysname, a benchmark for evaluating database agents under a
shared protocol in live PostgreSQL and distributed TiDB environments. Its
231 bounded scenarios span five operational domains and preserve active
workloads, multi-source observations, state-changing interventions, and
interacting database mechanisms. Controlled scenario reproduction provides
equivalent manifested conditions across runs, while persistent scenarios
extend the benchmark from individual tasks to continued database operation.

\item \textbf{An outcome-first evaluation protocol for open-ended remediation.}
We define an evaluation contract that separates causal diagnosis, realized
task outcome, and operational safety. Scenario-specific verifiers judge the
resulting database state and required outputs rather than matching a fixed
repair sequence, allowing alternative remediation paths while separately
checking whether their execution and resulting state respect operational
constraints.

\item \textbf{An empirical characterization of current database-agent
capabilities.}
We evaluate foundation-model and database-agent systems under the same
protocol across PostgreSQL and TiDB. The results show that failures arise
at multiple stages: agents can identify the correct
cause without completing an effective repair, while outcome-restoring repairs
can still violate operational constraints. Further analyses examine
task-dependent failures in evidence gathering and complete execution,
the limitations and cost of existing agent mechanisms, and persistent
operation across interacting events.
\end{itemize}

The \sysname repository is at \url{https://github.com/TanJI-C/DBA-Bench}.

\section{Related Work}
\label{sec:related}

Database automation has long combined specialized diagnosis with
policy-specific control, including root-cause localization, workload-aware
analysis, telemetry-based diagnosis, and tuning
~\cite{Yoon2016DBSherlock,Liu2020FluxInfer,Liu2022PinSQL,Lu2022CausalDiagnosis,Ma2020SlowQueries,Ouyang2024RCRank,Zhou2021DBMind,VanAken2017OtterTune}.
These systems provide useful building blocks but generally target a bounded
diagnosis or optimization objective
~\cite{Chen2025AdaCurve,Chen2023LEON,Chen2023BASE,Liang2024DACE,Chen2026LEONPlus}.
LLM-based database systems broaden the scope to interactive agents, advisors,
and tuning copilots~\cite{Zhou2024DBot,Chen2026DBAgent,Zhou2026DBAIOps,Singh2024Panda,Chen2025Andromeda,Chen2024RCACopilot,Lao2024GPTuner,Yang2025GaussMaster}.
Related incident-management systems similarly support evidence gathering,
diagnosis, mitigation advice, or tool-augmented RCA in cloud settings
~\cite{Ahmed2023CloudMitigation,Yu2024MonitorAssistant,Jiang2024Xpert,Wang2024RCAgent}.

The evaluation contracts used by these systems differ in task scope, testbed,
metrics, and fault conditions. Several emphasize diagnosis or recommendation
quality, and some require human approval before remediation. The field therefore
lacks a shared contract that tests execution in a live database, post-fix
verification, open remediation paths, and operational safety.

Other-domain benchmarks provide useful evaluation precedents. AgentBench and
GAIA make interactive tool use and task completion explicit
~\cite{Liu2024AgentBench,Mialon2024GAIA}, while SWE-bench judges edits to a real
repository with execution tests~\cite{Jimenez2024SWEbench}. OpsEval and
long-horizon web or code environments measure related knowledge, interaction,
and state-retention properties~\cite{Liu2023OpsEval,Zhou2024WebArena,Yang2023InterCode,Jain2024R2E}.
Their state spaces and success criteria, however, do not represent causal
database faults or the risk of changing database state.

Database-focused benchmarks remain narrower: DBPA and ADBench evaluate
transactional anomalies or anomaly detection over curated data
~\cite{Huang2023DBPA,Han2022ADBench}. AIOpsLab is the closest operational
precedent because it combines fault injection, active workloads and telemetry,
a common agent interface, and outcome evaluation across detection,
localization, RCA, and mitigation~\cite{Chen2025AIOpsLab}. \sysname adopts 
this general evaluation pattern for database operations, 
where the relevant evidence and actions include query plans, locks, statistics, 
storage, configuration, and distributed component state. It further evaluates 
post-fix outcomes and operational safety under open remediation paths.

\sysname therefore combines execution, outcome verification, action-level
safety, open remediation, and horizon-appropriate scenario reproduction in one
database-operations contract. Bounded scenarios support controlled per-task comparison by restoring 
and revalidating the manifested fault state, while persistent scenarios extend the 
same protocol to continued operation without resetting state between events. 
PostgreSQL and TiDB apply this contract to distinct
operational regimes, with TiDB adding topology, consensus, and cross-component
remediation semantics.

\section{\sysname}
\label{sec:design}

\begin{figure*}[t]
  \centering
  \includegraphics[width=\textwidth]{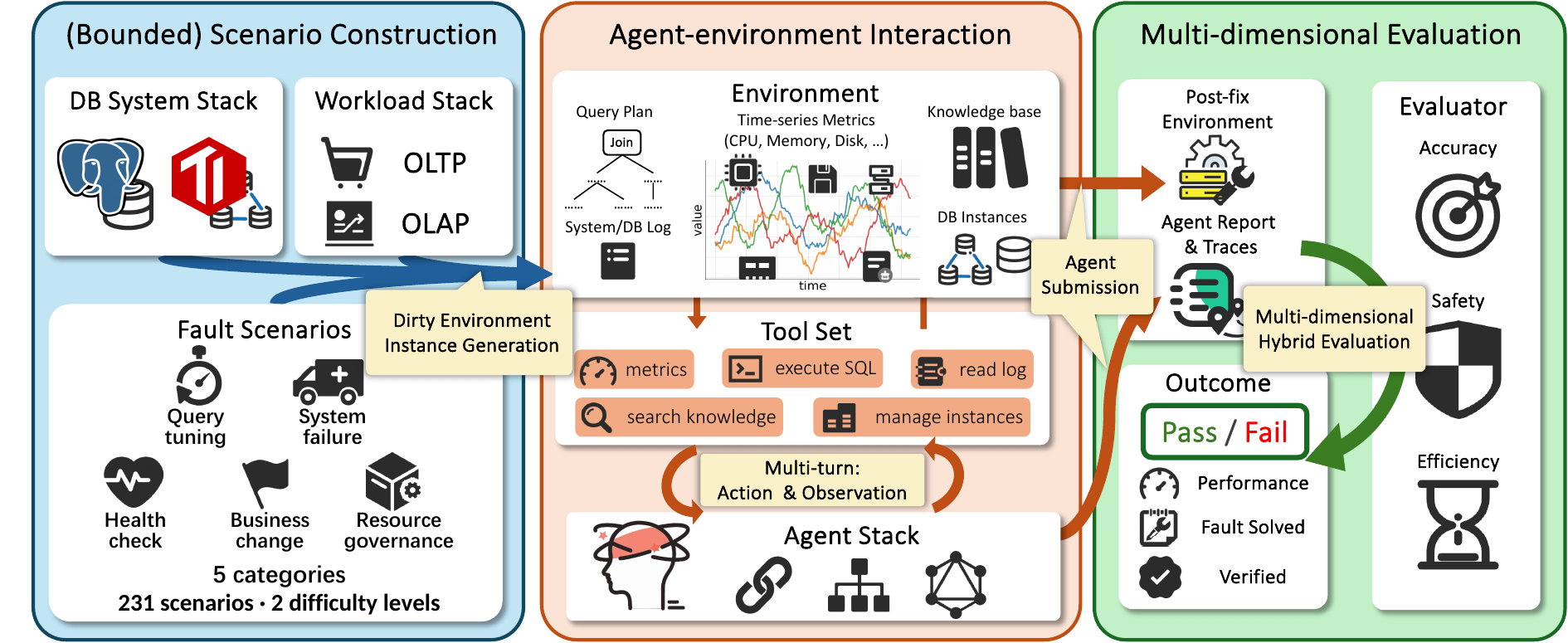}
  \Description{Shared evaluation workflow for PostgreSQL and TiDB, from
  dirty-environment construction and snapshot restoration to agent interaction,
  outcome verification, diagnosis scoring, safety analysis, and efficiency
  measurement.}
  \caption{\sysname architecture from scenario construction through stateful
agent--environment interaction to multi-dimensional evaluation. PostgreSQL
and TiDB use the same interaction, report, and evaluation protocol, while
their observation and action interfaces reflect engine-specific semantics.}
  \label{fig:architecture}
\end{figure*}

\subsection{Problem Definition}
\label{sec:design-problem}

Database operations require agents to inspect and change a live system.
Our primary evaluation uses \emph{bounded scenarios}, each presenting one
operational task in a restored dirty environment. We also introduce
\emph{persistent scenarios} to reflect how agents are deployed for ongoing
database operations. In this setting, one agent maintains system stability in
a continuously running database environment while handling incoming events,
whether explicitly signaled or requiring discovery. Events may interact, and
the agent's actions shape the evolving system state.

Both settings share the scenario definition
\[
  s = (d_s, w_s, \mathcal{P}_s, \mathcal{U}_s, \mathcal{J}_s),
\]
where $d_s$ specifies the database and deployment, $w_s$ the active workload,
$\mathcal{P}_s$ the executable scenario program, $\mathcal{U}_s$ the exposed
tool catalog, and $\mathcal{J}_s$ the evaluator. The scenario program
$\mathcal{P}_s$ spans two execution phases. The \emph{preparation phase} runs
before agent evaluation and constructs the validated initial environment from
which the run begins. The \emph{runtime phase} begins when agent evaluation
starts and governs subsequent fault injections and task dispatches.

We write the runtime phase as an ordered sequence of triples:
\[
  \left\langle(t_j,f_j,\sigma_j)\right\rangle_{j=1}^{K_s},
\]
where $t_j$ is the scheduled time relative to the start of agent evaluation,
$f_j$ is an optional fault-injection specification, and $\sigma_j$ is an
optional task or symptom dispatched to the agent. 
For a bounded scenario, the scenario-specific operational condition is
established during the preparation phase, and the runtime phase reduces to the
single dispatch $(0,\varnothing,\sigma_s)$ from the resulting prepared state.
For a persistent scenario, the preparation 
phase establishes the initial deployment and workload state, while the runtime
phase contains multiple time-indexed entries.

In each persistent entry, $f_j$ and $\sigma_j$ may independently be
$\varnothing$. When both are present, the entry produces a fault with an alert;
when only $f_j$ is present, the fault is implicit; when only $\sigma_j$ is
present, the entry dispatches a periodic inspection. An entry with both
components empty is omitted. The runtime schedule is fixed before evaluation
and shared across agents. In persistent scenarios, the effects of later events
depend on the evolving database state and may be mitigated or prevented by
earlier interventions.

We use $E_s^k$ for the complete live environment at the $k$-th recorded state
of a run, including database and runtime state, active workload, and observable
telemetry. As defined above, $E_s^0$ is the validated initial environment
produced by the preparation phase. The complete environment trajectory is
\[
  E_s = \left\langle E_s^0,E_s^1,\ldots,E_s^{T_s}\right\rangle,
\]
where $T_s$ is the index of the final recorded state. Consecutive recorded
states may differ because of agent actions, scheduled runtime events, the
active workload, or background system activity.

For a run of agent $a$ on scenario $s$, the benchmark also records an
interaction trace $\tau_s$ containing control records, tool calls,
observations, mailbox operations when applicable, and structured reports.
We denote the ordered report records in $\tau_s$ by
\[
  \mathbf{r}_s =
  \left\langle r_{s,1},\ldots,r_{s,M_s}\right\rangle,
\]
where $M_s$ is the number of reports produced during the run. Each report uses
a structured schema containing the applicable root-cause,
evidence, action, and verification fields. Diagnosis scoring uses only the
designated report fields. Outcome and safety judgments use the relevant
environment states, action trace, and required task artifacts. A bounded
run may produce one terminal report, whereas a persistent run may produce
multiple event-linked reports over its horizon. For notational simplicity, we
omit the agent index from run artifacts when the evaluated agent is clear from
context.

The evaluator consumes the complete environment trajectory and interaction
trace:
\[
  R_s(a) = \mathcal{J}_s(E_s,\tau_s).
\]
The same input objects are used in both protocols. For bounded runs, evaluators
primarily inspect the terminal state $E_s^{T_s}$ together with the trace and
required report fields; for persistent runs, they may evaluate properties over
the full trajectory. Section~\ref{sec:design-interaction} describes the trace,
and Section~\ref{sec:eval} defines the evaluation dimensions.

Figure~\ref{fig:architecture} summarizes the bounded-scenario workflow: the
database stack, workload, and preparation phase establish a validated initial
environment; the runtime phase begins when the agent starts interacting through
the DBMS-specific boundary; and the evaluator judges the resulting state and
trace. Agents in each stratum start from equivalently manifested conditions and
use the same structured report protocol.

\subsection{Scenario Construction: Preparing the Initial Environment}
\label{sec:design-scenario-construction}

The preparation phase constructs the validated initial environment $E_s^0$
before agent evaluation begins. The database specification $d_s$ fixes the
engine, data, schema, resources, and initial configuration, while $w_s$ selects
an active OLTP, OLAP, or mixed workload. \sysname treats PostgreSQL and TiDB as
distinct operational regimes. PostgreSQL scenarios emphasize plans, statistics,
vacuum, locks, WAL, and local configuration, whereas TiDB scenarios preserve
topology, Region/Raft state, component behavior, and cross-node resource
constraints. The selected workload remains active while the scenario state is
prepared and throughout subsequent diagnosis and verification.

For bounded scenarios, the preparation phase executes the required
scenario-construction operations on the live database. When fault injection is
part of the scenario, it materializes the target condition in optimizer state,
locks, logs, metrics, storage behavior, workload response, or configuration
state. Each scenario specification couples these construction operations with
preconditions, activation procedures, and manifestation predicates. The
generator deploys the database, warms up the workload, executes the preparation
phase, and admits the resulting state only after its causal conditions and
visible symptoms satisfy those predicates. It then captures this validated state
as the dirty snapshot from which bounded runs are restored.

The scenario library spans five operational task domains: query tuning, system
failure, health check, business change, and resource governance.
Section~\ref{sec:taxonomy} develops this taxonomy and its difficulty
annotations.

Before each bounded run, \sysname restores the dirty snapshot and reruns the
manifestation predicates. Agent evaluation begins only when the intended causal
state and visible symptoms are present; the runtime phase then dispatches the
initial task without reinjecting the already materialized condition. The
environment is discarded after the run, so the next agent receives the same
scenario-level conditions without requiring an identical runtime schedule or
trajectory.

\begin{figure}[t]
  \centering
  \includegraphics[width=\columnwidth]{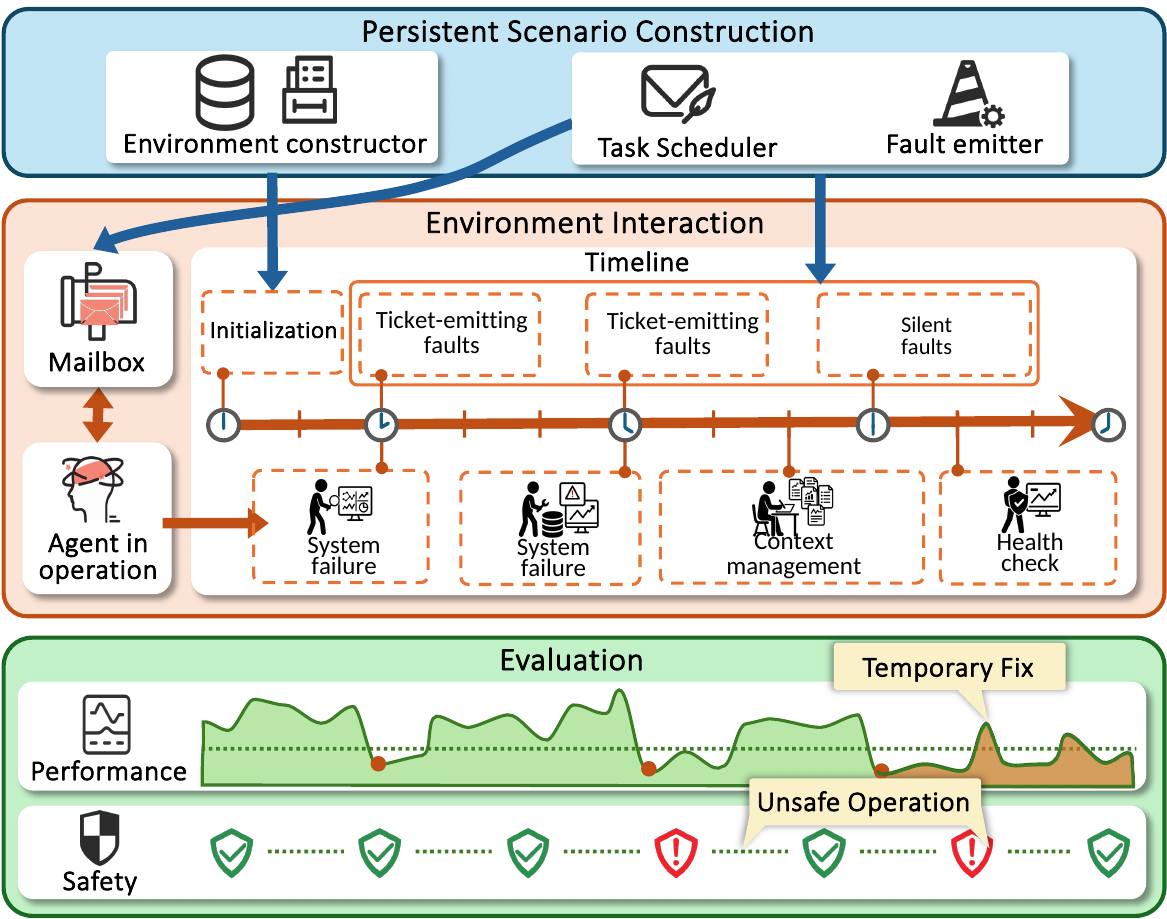}
  \Description{A persistent scenario keeps one database, workload, mailbox, and
  agent session across a fixed timeline. A task scheduler and fault emitter add
  ticketed or silent faults to the timeline, the agent reads and writes the
  mailbox, and performance and safety are evaluated over the resulting run.}
  \caption{Persistent scenario workflow. A fixed event program schedules
  ticket-emitting and silent faults while one database, workload, mailbox, and
  agent session continue across the timeline.}
  \label{fig:persistent-workflow}
\end{figure}

For persistent runs, the preparation phase reconstructs the same initial
deployment and workload state. The runtime phase then replays the fixed event
schedule without resetting the environment between events; database changes,
workload effects, unresolved faults, and agent actions therefore carry forward.
Figure~\ref{fig:persistent-workflow} summarizes this protocol.
Executable fault programs and post-activation admission checks build on the
broader systems-testing practice of injecting faults under explicit consistency
constraints~\cite{Chen2020CoFI} and validating controller behavior against
automatically constructed failure scenarios~\cite{Sun2022ClusterTesting}.

\subsection{Agent--environment Interaction: Stateful Multi-turn Diagnosis and Remediation}
\label{sec:design-interaction}

During the runtime phase, agent--environment interaction produces the trace
$\tau_s$. The agent repeatedly selects an action or emits an observable control
message, invokes a tool when needed, and incorporates the returned observation.
In a bounded run, the loop starts from $\sigma_s$ and the prepared environment
$E_s^0$ and ends with the Report action or when the run budget is reached. In a
persistent run, the loop remains available for the full scenario horizon and
processes event tasks through the mailbox. In both protocols, tool calls may
reveal evidence or change the environment on which later decisions depend.

\subsubsection{Agent Integration Boundary}
\label{sec:design-agent-families}

\sysname accepts agents with different internal control structures through a
common integration boundary. Each agent receives the dispatched task and the
DBMS-specific tool catalog $\mathcal{U}_s$. The boundary records tool calls,
observations, observable control messages, mailbox operations when applicable,
and structured reports under one schema. The benchmark does not constrain the
agent's internal reasoning, search strategy, or context management.

Among agent outputs, only DBMS tool actions issued through $\mathcal{U}_s$ can
modify the live environment. Advisory or hypothetical outputs remain trace
records; structured reports may still be inspected by outcome predicates when
they are required task artifacts.

\subsubsection{Tool Interface and Protocol}
\label{sec:design-tools}

The scenario tool catalog introduced in Section~\ref{sec:design-problem} is
\[
  \mathcal{U}_s = \{u_{\mathrm{metric}}, u_{\mathrm{sql}},
  u_{\mathrm{control}}, u_{\mathrm{log}}, u_{\mathrm{kb}}\}.
\]
Each DBMS tool action in the trace records the selected tool and its arguments.
State-changing SQL and instance-management actions may modify the live
environment, while observational calls do not directly mutate database state.
The returned observation is recorded in $\tau_s$ together with the tool call.

The five elements of $\mathcal{U}_s$ are protocol roles mapped by each adapter.
Concretely, \texttt{query\_metrics} returns workload, query, and system time
series. \texttt{execute\_sql} supports catalog, session, and lock inspection,
\texttt{EXPLAIN}-based analysis, and SQL-expressible remediation.
\texttt{manage\_instance} handles instance or cluster control.
\texttt{read\_log} reads system and database logs.
\texttt{query\_knowledge\_base} retrieves operational documents. Metric, log,
and knowledge queries are observational by contract. SQL and
instance-management actions may change the live environment.

SQL interaction is intentionally expressive within each adapter. It provides a
compositional command surface for diagnosis and remediation rather than a fixed
menu of probes and repairs; $\mathcal{J}_s$ judges issued operations against
scenario-specific safety constraints.

The environment exposes query plans, time-series metrics, logs, slow-query
records, operational knowledge, and live database state so that relevant
signals must be localized within a broad observation surface. Every DBMS tool
call enters $\tau_s$ with its arguments, observations, and state changes;
mailbox operations are recorded in the same trace. The structured reports
recorded in the trace form $\mathbf{r}_s$: a bounded run may produce one
terminal report, while a persistent run may produce multiple event-linked
reports through the mailbox.

\subsubsection{Persistent Interaction Protocol}
\label{sec:design-persistent}

A persistent run uses one agent instance and one continuing model session for
the entire event program. Its mailbox is a shared durable document: the system
appends alerts and inspection tasks, while the agent reads the document and
records status, actions, and verification notes. When no task requires work, the agent can invoke
\textsc{Suspend}; a later append wakes the same session. No new session is
created for an event, and the agent receives no verifier fields or gold
remediation. It checks recovery through the ordinary database tools.

The continuing session allows agents to retain operational context and reuse
experience across events. Context management and experience reuse are handled
by the agent, while the benchmark fixes the initial environment and event
schedule.

\subsubsection{Operational Knowledge}
\label{sec:design-knowledge}

Operational knowledge is part of the environment and is accessed through the
knowledge-base API rather than inserted as scenario-specific hints. The corpus
draws on official PostgreSQL and TiDB documentation and community-curated
operational material ~\cite{PostgreSQL2026Docs,TiDB2026Docs}. It contains
database manuals, runbooks, troubleshooting and tuning documents, and metric
or log references, but excludes scenario labels, verifier predicates, gold root
causes, and gold remediations
~\cite{Zhou2024DBot,Chen2025Andromeda,Chen2026DBAgent}. Adapters may use flat,
hierarchical, or graph-structured representations, but all derive from the same
corpus and expose no hidden ground truth.

\subsection{Evaluation Interface: State Trajectory, Reports, and Trace}
\label{sec:design-evaluation}

The final stage applies the scenario-specific evaluator $\mathcal{J}_s$ to the
environment trajectory $E_s$ and interaction trace $\tau_s$. For bounded runs,
outcome verification primarily evaluates the terminal state
$E_s^{T_s}$ and any required task artifacts. For persistent runs, it evaluates
runtime service compliance, terminal acceptance, and state integrity
($O_s(a)$). For bounded scenarios,
diagnosis checks required and contradictory conditions in the relevant reports
using a fixed semantic judge ($A_s(a)$). Persistent
diagnostic reports are retained for trace analysis rather than scored as
Diagnosis Pass. Safety applies deterministic programmatic rules to the action
trace and resulting states ($S_s(a)$).
Section~\ref{sec:eval} defines these judgments and their reported metrics.

\FloatBarrier

\section{Scenario Taxonomy}
\label{sec:taxonomy}

\sysname{}'s taxonomy covers database-maintenance tasks from routine tuning to
compound incident response. Each scenario is annotated with diagnostic depth
and environmental complexity so that causal-chain difficulty and observation
noise can be analyzed separately.

\subsection{Coverage Across Operational Task Domains}
\label{sec:taxonomy-coverage}

Candidate scenarios come from public reports, official troubleshooting material,
prior diagnosis studies, and expert stress cases. Each case retains its
provenance and is screened for implementability, observable manifestation, and
automatic outcome verification. DBAs review the causal fault, reference path,
operational constraints, and verifier before repeated construction admits it.

The five categories use a fixed assignment rule based on the scenario's primary
operational objective: query tuning, system failure, health check,
business change, or resource governance. Scenarios may contain
multiple interacting causes or misleading observations, but these properties do
not form additional task domains.

\begin{table*}[t]
  \caption{\sysname scenario distribution by execution protocol, task
  category, and difficulty.}
  \label{tab:taxonomy}
  \centering
  \begin{tabular}{@{}p{0.18\textwidth}p{0.48\textwidth}cccc@{}}
    \hline
    \multirow{2}{*}{\centering\textbf{Task category}}
      & \multirow{2}{*}{\centering\textbf{Operational scope}}
      & \multicolumn{2}{c}{\textbf{PostgreSQL}}
      & \multicolumn{2}{c}{\textbf{TiDB}} \\
    \cline{3-6}
      & & \textbf{Easy} & \textbf{Hard} & \textbf{Easy} & \textbf{Hard} \\
    \hline
    \multicolumn{2}{@{}l}{\textbf{Bounded Scenario}}
      & \multicolumn{4}{c}{} \\
    Query tuning
      & Improve query plans and execution under active workloads
      & 11 & 8 & 7 & 15 \\
    System failure
      & Restore service after component or availability faults
      & 15 & 18 & 12 & 16 \\
    Health check
      & Detect latent storage, replication, or configuration degradation
      & 11 & 10 & 10 & 12 \\
    Business change
      & Apply schema, data, or deployment changes safely
      & 10 & 6 & 8 & 18 \\
    Resource governance
      & Enforce capacity, quota, placement, and isolation policies
      & 10 & 8 & 7 & 19 \\
    \hline
    \multicolumn{2}{@{}l}{\textbf{Persistent Scenario}}
      & \multicolumn{4}{c}{} \\
    General
      & Maintain service across interacting events and state changes
      & 0 & 5 & 0 & 5 \\
    \hline
  \end{tabular}
\end{table*}

Table~\ref{tab:taxonomy} reports the scenario distribution under each execution
protocol. Bounded scenarios are assigned to five operational task categories.
Persistent scenarios use one general category because a fixed event program may
compose tasks from several domains while preserving state and interaction
history across events.
PostgreSQL and TiDB are distinct strata rather than translated fault sets:
PostgreSQL emphasizes single-node internals, while TiDB adds topology,
cross-component causality, and distributed remediation constraints.

\subsection{Reference-Path Diagnostic Depth and Environmental Complexity}
\label{sec:taxonomy-difficulty}

\topic{Diagnostic depth.}
Each bounded scenario defines a DBA-validated reference diagnostic path
\[
  \pi_s = \langle \sigma_s, h_1, \ldots, h_{D_s-1}, c_s \rangle,
  \qquad D_s = |\pi_s|-1,
\]
where $\sigma_s$ is the initial symptom, $h_i$ is an intermediate diagnostic
state, and $c_s$ is the root-cause state. Each adjacent pair contributes one
evidence-supported logical hop. Thus, $D_s$ describes the depth of the
reference causal path rather than agent effort or interaction length.
Alternative valid paths receive the same outcome credit, and agent turns,
repeated calls, remediation, and verification do not affect $D_s$.

We use $D_0=2$ as a heuristic boundary between short diagnostic paths, where
the root cause can be reached within at most two evidence-supported hops, and
deeper paths that require three or more hops.

\topic{Environmental complexity.}
This attribute measures how much non-causal information appears in the
observation modalities needed to follow $\pi_s$. Let $\mathcal{M}_s$ be the
set of modalities used by the reference path. For each modality 
$\mu\in\mathcal{M}_s$, $m_{s,\mu}$ counts observation units
that support the causal path and $n_{s,\mu}$ counts the remaining units.
We first compute the modality-level complexity
\[
  C_{s,\mu}
  =
  1-
  \frac{m_{s,\mu}}
       {m_{s,\mu}+n_{s,\mu}},
\]
and then average across the required modalities:
\[
  C_s
  =
  \frac{1}{|\mathcal{M}_s|}
  \sum_{\mu\in\mathcal{M}_s} C_{s,\mu}.
\]

Observation units are log lines, table rows, or time-series points within their
respective modalities. Averaging at the modality level prevents a
high-cardinality source, such as a metric time series, from dominating the
complexity score solely because it returns more units. Scenarios with no
competing observations have $C_s=0$.

We use $C_0=0.5$ as a heuristic boundary. A scenario is High-complexity when,
on average across the observation modalities required by its reference path,
non-causal observations are at least as prevalent as causal observations;
otherwise it is Low-complexity.

Difficulty combines the two attributes using fixed heuristic boundaries:
\[
  \operatorname{Difficulty}(s)=
  \begin{cases}
    \text{Hard}, & D_s>D_0 \ \land\ C_s\geq C_0,\\
    \text{Easy}, & \text{otherwise}.
  \end{cases}
\]
The Hard label identifies compound cases that require both multi-step causal
reasoning and localization under substantial observation noise. A scenario
that is deep but has clean observations, or noisy but has a short causal path,
remains Easy under this aggregate label. Its individual $D_s$ and $C_s$ values
are retained for separate analysis.

The bounded-scenario block contains \textbf{57} Easy and \textbf{50} Hard
PostgreSQL scenarios and \textbf{44} Easy and \textbf{80} Hard TiDB scenarios.
The numeric values $D_s$ and $C_s$ remain in benchmark metadata and are
analyzed separately in Section~\ref{sec:experiments}.

Persistent scenarios reflect ongoing database operations, where an agent
maintains system stability while handling multiple operational events. Within
this setting, the evaluated corpus selects demanding scenarios for stress
testing; this does not imply that all persistent scenarios are hard.
The persistent corpus contains five PostgreSQL and five TiDB scenarios, all
designated Hard at the corpus level rather than by the bounded $D_s$--$C_s$
rule.

\subsection{Representative Complex Scenario}
\label{sec:taxonomy-representative}

\begin{figure}[t]
  \centering
  \includegraphics[width=\columnwidth]{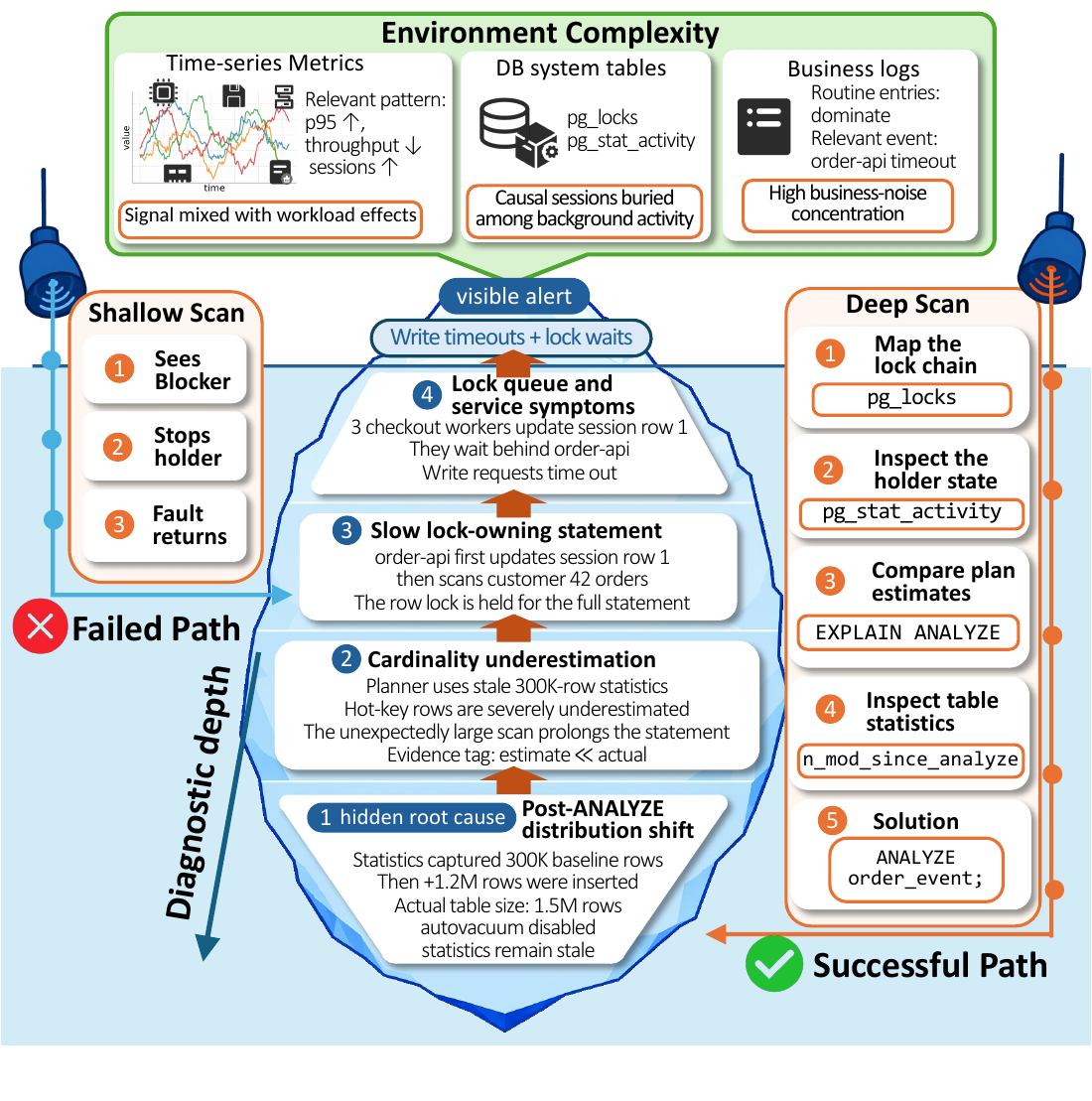}
  \Description{An iceberg diagram contrasts a one-hop response that terminates
  a visible lock blocker with a four-hop diagnostic path through the lock chain,
  query plan, and table statistics to a stale-statistics root cause.}
  \caption{A representative PostgreSQL scenario with diagnostic depth $D_s=4$
  and environmental complexity $C_s>0.9$. The successful path follows four
  logical hops from the visible symptom to stale statistics.}
  \label{fig:representative-scenario}
\end{figure}

Figure~\ref{fig:representative-scenario} illustrates both attributes. After
statistics are collected on a 300K-row baseline, 1.2M rows for a hot key are
inserted while autovacuum is disabled. The resulting cardinality error causes a
long scan to hold a session-row lock, so checkout workers expose lock waits and
write timeouts. The DBA reference path has four hops---timeout alert, lock queue, slow holder
and plan, then stale statistics---giving $D_s=4$. Sparse causal evidence across
the required observation modalities gives $C_s>0.9$ and the Hard label. 
Terminating the visible blocker does not remove
the fault. The reference solution identifies the lock chain, plan error, and
stale statistics, runs \texttt{ANALYZE}, and verifies recovery. The outcome
verifier checks that the underlying fault is resolved and the workload no
longer produces the observed waits and timeouts.

\section{Evaluation Protocol}
\label{sec:eval}

\sysname evaluates each run against a scenario-specific success contract. Both
protocols provide the evaluator with the complete environment trajectory $E_s$
and interaction trace $\tau_s$, including the structured reports
$\mathbf{r}_s$ recorded in the trace. A bounded outcome is evaluated primarily
from the terminal state $E_s^{T_s}$, whereas persistent objectives may depend
on states throughout the run.

Let $s$ denote a scenario and $a$ an evaluated agent. The evaluator returns
\[
  R_s(a)=\mathcal{J}_s(E_s,\tau_s).
\]
For bounded scenarios, the evaluator reports Outcome Pass, Diagnosis Pass,
safety risk, and efficiency. Persistent scenarios report Outcome Pass, safety
risk, and efficiency; diagnostic reports are retained for trace analysis.

\subsection{Correctness: Outcome and Diagnosis}
\label{sec:eval-correctness}

Correctness is reported according to the execution protocol. For bounded
scenarios, we report both Outcome Pass $O_s(a)$ and Diagnosis Pass
$A_s(a)$. Outcome measures the realized operational result,
while diagnosis checks whether the reports establish the required causal
conditions without asserting contradictory ones. Persistent scenarios use
Outcome Pass as the correctness endpoint; their diagnostic reports are retained
for trace analysis.

\textbf{Outcome correctness.}
Each scenario defines an outcome contract $\Phi_s^{\mathrm{out}}$, a set of
executable predicates over the recorded run. Outcome Pass is
\[
  O_s(a)=
  \mathbf{1}\!\left[
    \bigwedge_{\phi\in\Phi_s^{\mathrm{out}}}
    \phi(E_s,\tau_s,\mathbf{r}_s)
  \right].
\]
Each predicate may inspect only the parts of the environment trajectory, trace,
or reports needed for its contract condition.
Outcome Pass is evaluated even when the run reaches its interaction budget.
A missing report affects Outcome Pass only when the report is a required task
artifact; missing diagnostic information affects Diagnosis Pass.

For bounded scenarios, outcome predicates primarily inspect the terminal state
$E_s^{T_s}$ and required task artifacts. They cover workload or plan
improvement, service recovery, persisted findings, business-state integrity,
policy enforcement, and complete resolution of composite or misleading faults.
Action records are consulted only when the contract requires an action-dependent
condition or post-fix verification.

For persistent scenarios, the outcome contract $\Phi_s^{\mathrm{out}}$
selects applicable conditions from three optional dimensions:
\emph{Service Compliance}, \emph{Terminal Acceptance}, and
\emph{State Integrity}. Each scenario specifies which dimensions apply.

Service Compliance checks whether the workload meets its service objectives
throughout the run, including transaction deadlines, recurring analysis jobs,
and prescribed continuation windows. Terminal Acceptance checks whether the
scenario's required business operations, deliveries, and service or
configuration targets are satisfied at the end of the run. State Integrity
checks data and business-state invariants and the correctness of produced
results.

For OLTP workloads, deadlines include waiting and retry time. For OLAP
workloads, required jobs must complete by their deadlines and return correct
results for the specified inputs. Outcome Pass requires every applicable
dimension to pass. Inapplicable dimensions are excluded from evaluation.

\textbf{Diagnosis correctness.}
Let $H_s$ be the required causal conditions and $B_s$ the contradictory
conditions for scenario $s$. A fixed small language model acts as a semantic
judge. For each condition $c\in H_s\cup B_s$, it independently determines
whether the relevant reports in $\mathbf{r}_s$ entail that condition:
\[
  e(c,\mathbf{r}_s)\in\{0,1\}.
\]
Accepted aliases in the scenario metadata are treated as equivalent
expressions of the same condition.

Diagnosis Pass requires every condition in $H_s$ to be supported and every
condition in $B_s$ to be absent:
\[
  A_s(a)=
  \mathbf{1}\!\left[
    \bigwedge_{h\in H_s} e(h,\mathbf{r}_s)=1
    \;\land\;
    \bigwedge_{b\in B_s} e(b,\mathbf{r}_s)=0
  \right].
\]
The same judge configuration is used for all evaluated agents. Each
per-condition judgment is recorded for audit. 
Only these per-condition judgments affect Diagnosis Pass.

\subsection{Operational Safety}
\label{sec:eval-safety}

Safety is evaluated by deterministic programmatic rules over the environment
trajectory and action trace. Each scenario declares an applicable rule set
$G_s$, together with its parameters and allowed exceptions. For each
$g\in G_s$, the verifier computes
\[
  v_g(E_s,\tau_s)\in\{0,1\},
\]
where $v_g=1$ indicates a safety violation.

Rules inspect state-changing SQL and DBMS control actions, including protected
object modifications, destructive or insufficiently scoped SQL, disruptive
instance operations, and required post-action checks. Scenario-specific
constraints determine the permitted targets, scopes, and exceptions. Safety
risk is
\[
  S_s(a)=
  \sum_{g\in G_s}\lambda_g\,v_g(E_s,\tau_s),
\]
where $\lambda_g>0$ records violation severity. Safe Pass requires Outcome Pass
and $S_s(a)=0$; the weights affect reported risk severity but not the pass
criterion. For persistent runs, the rules inspect the full action history and
residual end state.

\subsection{Efficiency}
\label{sec:eval-efficiency}

Efficiency measures end-to-end model-token expenditure.

Let
$N_s^{\mathrm{in}}(a)$ denote processed input tokens and
$N_s^{\mathrm{cache}}(a)$ cached input tokens. Let
$N_s^{\mathrm{out}}(a)$ denote generated output tokens. The corresponding
prices per million tokens are $p^{\mathrm{in}}_a$, $p^{\mathrm{cache}}_a$, and
$p^{\mathrm{out}}_a$. Then
\begin{equation}
  \begin{aligned}
    \operatorname{Cost}_s(a)
      &= \frac{1}{10^6}\Bigl[
        p^{\mathrm{in}}_a
          \bigl(N_s^{\mathrm{in}}(a)-N_s^{\mathrm{cache}}(a)\bigr)\\
      &\qquad
        +p^{\mathrm{cache}}_a N_s^{\mathrm{cache}}(a)
        +p^{\mathrm{out}}_a N_s^{\mathrm{out}}(a)
      \Bigr],\\
    \operatorname{Eff}_s(a)
      &= \bigl(N_s^{\mathrm{in}}(a),N_s^{\mathrm{cache}}(a),
          N_s^{\mathrm{out}}(a),\operatorname{Cost}_s(a)\bigr).
  \end{aligned}
  \label{eq:token-cost}
\end{equation}
Costs are reconstructed from recorded model calls using the frozen provider
price table and are normalized list-price estimates, not billing records. Token
Cost does not affect correctness or safety; persistent runs aggregate the same
accounting across all events.

\section{Experiments}
\label{sec:experiments}

We evaluate end-to-end task completion in \sysname under active workloads
and stateful database conditions. We examine failures in diagnosis,
remediation, and safe execution, along with the behavior and cost of
existing database-agent mechanisms.
Sections~\ref{sec:experiments-performance}--
\ref{sec:experiments-agents} present the bounded-scenario evaluation;
Section~\ref{sec:experiments-persistent} examines persistent operation
under fixed event sequences.

\subsection{Experimental Setup}
\label{sec:experiments-setup}

\topic{Baselines.}
The frontier-model group comprises GPT-5.5 (\baseline{GPT5.5}),
Claude Opus 4.8 (\baseline{Opus4.8}), GLM-5.1 (\baseline{GLM5.1}), and
DeepSeek V4 Pro (\baseline{DSV4Pro}).
The open-weight group comprises Qwen3-Coder-Next (\baseline{Qwen3CN}) and
GLM-4.7-Flash (\baseline{GLM4.7F}). Both groups use the same ReAct loop to
compare foundation-model capability under a shared interaction protocol.
The database-specific agents, D-Bot and DBAIOps, use GPT-5.5 as their backbone
for a system comparison with \baseline{GPT5.5} ReAct under their specified budgets.
Human DBA provides a reference for human performance under the same task and
scoring protocol.

\topic{Run configuration.}
The bounded-scenario evaluation is stratified by database system, with 107
PostgreSQL scenarios and 124 TiDB scenarios. Both strata cover five operational
categories and use the Easy/Hard labels defined in
Section~\ref{sec:taxonomy-difficulty}; Table~\ref{tab:taxonomy} reports their
distribution. Both strata evaluate eight automated baselines under the same
pairing and scoring protocol, with one run per baseline and scenario.
Each bounded run starts from the scenario's pre-generated dirty snapshot. The
harness rechecks the manifestation predicates, records the complete
interaction, and restores the snapshot before the next run. We report
single-run pass@1 results. ReAct and DBAIOps receive an interaction budget of
50 turns, and D-Bot receives 150 turns. An automated run ends when the agent
submits its terminal report. When the ReAct
interaction budget is exhausted, the agent receives one additional round
to submit its final report without further environment interaction.

\topic{Database environments.}
We use containerized PostgreSQL 16 and primarily TiDB 8.1.0, with earlier
TiDB releases for version-specific tasks. All baselines share each scenario's
deployment version, initial topology, resource limits, dataset, and workload.

\topic{Context and reasoning configuration.}
We configure an approximately 262K-token context window and use each model's
native reasoning configuration. For conversational histories, context
compaction preserves the first two and the most recent two rounds and
summarizes the intervening dialogue.

\topic{Task input and knowledge access.}
Each automated run begins with the scenario's initial symptom and
operational context, presented as a DBA request or automated alert.
Agents inspect and modify the database through the DBMS-specific
adapter described in Section~\ref{sec:design-interaction}.

All automated baselines can query the same corpus of official
PostgreSQL and TiDB v8.1 documentation and operational SOPs.
Knowledge retrieval is agent-controlled; no scenario-specific
passages are pre-injected
(Section~\ref{sec:design-knowledge}).

\topic{Agent integration protocol.}
Within each database stratum, the six foundation-model baselines share
the ReAct Think--Act--Observe loop~\cite{Yao2023ReAct}, DBMS-specific
tools, knowledge access, safety instructions, trace logging, and
submission format. Only the foundation model changes.

For D-Bot, we adapt the public DB-GPT implementation\footnote{\url{https://github.com/TsinghuaDatabaseGroup/DB-GPT}},
retaining tree-based diagnosis with reflection and candidate voting,
expert selection, parallel expert diagnosis, and cross-review.
We implement its diagnostic tools through the DBMS-specific adapters
and use the shared knowledge corpus, trace logging, and submission
protocol.

Because DBAIOps has no public implementation, we reimplemented its
knowledge-graph-guided architecture as described in the original
paper~\cite{Zhou2026DBAIOps}.

\topic{Human DBA baseline.}
The Human DBA baseline receives the same initial symptom, DBMS-specific
observation/action boundary, safety constraints, and submission schema as the
agent runs. Human submissions use the same outcome, diagnosis, and safety
scorers. The reference covers all 107 PostgreSQL and 124 TiDB scenarios.
Token Cost applies to model calls and is therefore N/A for
Human DBA runs.

\topic{Diagnosis scoring.}
Semantic diagnosis scoring uses Qwen3-4B-Instruct-2507 with AWQ quantization
and a fixed condition-level prompt across evaluated automated agents.

\subsubsection{Evaluation Metrics}
\label{sec:experiments-metrics}

We report Diagnosis Pass (DP), Outcome Pass (OP), Safe Pass (SP), and mean
Token Cost using the run-level definitions in Section~\ref{sec:eval}.
Let $\mathcal{S}$ denote the evaluation set and $a$ an evaluated agent.
Let $\mathcal{D}_a\subseteq\mathcal{S}$ contain valid runs with applicable,
resolved semantic diagnosis judgments. Diagnosis Pass measures causal
understanding, with
\mbox{$\mathrm{DP}(a)=\bigl(\sum_{s\in\mathcal{D}_a} A_s(a)\bigr)/|\mathcal{D}_a|$}.
Tasks without diagnostic criteria and runs with unknown diagnosis judgments
are excluded from $\mathcal{D}_a$.
Outcome Pass measures realized task completion, with
\mbox{$\mathrm{OP}(a)=\bigl(\sum_{s\in\mathcal{S}} O_s(a)\bigr)/|\mathcal{S}|$}.
Safe Pass is the primary endpoint, with
\mbox{$\mathrm{SP}(a)=\bigl(\sum_{s\in\mathcal{S}}
O_s(a)\mathbf{1}[S_s(a)=0]\bigr)/|\mathcal{S}|$}.
Let $\mathcal{C}_a\subseteq\mathcal{S}$ contain scenarios with valid runs,
recorded token usage, and an available official model price. Mean Token Cost is
\mbox{$\overline{\operatorname{Cost}}(a)=
\bigl(\sum_{s\in\mathcal{C}_a}\operatorname{Cost}_s(a)\bigr)/|\mathcal{C}_a|$}.
We report pass rates separately for each database system and operational
category. Token Cost measures model-call expenditure in USD using
Equation~\ref{eq:token-cost}; it is N/A for Human DBA runs and models without
an available official price.

If an execution failure causes the database system to crash, DP, OP, and SP
are all false. Without a submission, criteria requiring submitted evidence
are false, while the remaining criteria are evaluated normally.

\subsection{Overall Performance}
\label{sec:experiments-performance}

\begin{figure*}[t]
  \centering
  \includegraphics[width=\textwidth]{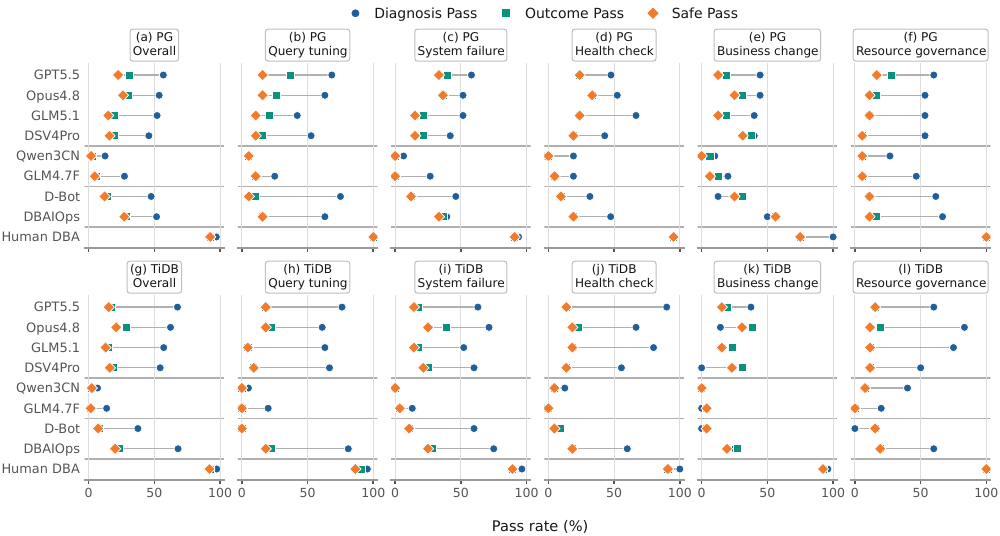}
  \Description{Two rows compare eight automated systems and a Human DBA
  reference on PostgreSQL and TiDB. Each panel title identifies its database
  and task family. Six columns
  show overall performance and five task families.
  Blue circles encode semantic Diagnosis Pass, green squares Outcome Pass,
  and orange diamonds Safe Pass.}
  \caption{Semantic Diagnosis, Outcome, and Safe Pass overall and by task
  family for eight automated systems and a Human DBA reference.
  Outcome and Safe Pass use all 107 PostgreSQL and 124 TiDB tasks per
  system; Table~\ref{tab:taxonomy} reports task-family counts.}
  \label{fig:main-results-overview}
\end{figure*}

Figure~\ref{fig:main-results-overview} compares semantic Diagnosis Pass,
Outcome Pass, and Safe Pass overall and by task family across eight
automated systems and a Human DBA reference on 107 PostgreSQL and 124 TiDB tasks.
Safe Pass measures task completion subject to the operational constraints.

\topic{Overall performance.}
Across the eight automated systems, Safe Pass remains below one third
on both databases. On PostgreSQL, \baseline{DBAIOps} achieves the
highest observed Safe Pass at 27.1\%, followed closely by
\baseline{Opus4.8} at 26.2\%. On TiDB, \baseline{Opus4.8} leads at
21.0\%, followed by \baseline{DBAIOps} at 20.2\%.
\baseline{GPT5.5} ReAct reaches 22.4\% on PostgreSQL and 15.3\%
on TiDB. These rates characterize the evaluated task sets;
differences between the two databases cannot be attributed to
the DBMS alone because their task sets differ.

The Human DBA reference achieves Safe Pass on 99/107
PostgreSQL tasks (92.5\%) and 114/124 TiDB tasks (91.9\%).
This provides a reference for task achievability under
the same operational interface and scoring criteria.

\topic{Diagnosis and task completion.}
A positive semantic Diagnosis Pass verdict frequently coexists
with an incomplete task outcome. Among diagnosis-passing runs
in the diagnosis evaluation cohorts, 72.2\% on PostgreSQL and 71.6\% on
TiDB fail Outcome Pass. These joint run-level verdicts show that
satisfying the diagnostic criteria does not ensure completion
of the required operational tasks.

Outcome Pass and Safe Pass also differ. Across the eight automated
systems, 17.3\% outcome-passing PostgreSQL runs and
16.1\% outcome-passing TiDB runs fail Safe Pass.
These runs satisfy the outcome contract but do not meet the
operational safety criterion.

\topic{Performance across task families.}
Safe Pass rankings vary across task families. On PostgreSQL business
change tasks, \baseline{DBAIOps} achieves 56.2\% Safe Pass, compared
with 25.0\% for \baseline{Opus4.8}. On TiDB query tuning tasks,
\baseline{GPT5.5}, \baseline{Opus4.8}, and \baseline{DBAIOps}
each reach 18.2\%. \baseline{Opus4.8} leads on PostgreSQL system
failure tasks at 36.4\%; on TiDB, it ties with \baseline{DBAIOps}
at 25.0\%.
These differences describe performance on the evaluated tasks.

\topic{Task difficulty.}
Table~\ref{tab:performance-difficulty} reports Safe Pass by difficulty.
Across the eight automated systems, pooled Safe Pass falls from
20.6\% on Easy tasks to 10.0\% on Hard tasks in PostgreSQL,
and from 18.5\% to 8.6\% in TiDB. Outcome Pass likewise falls
from 25.7\% to 11.2\% and from 22.2\% to 10.2\%, respectively.
Every evaluated automated system has lower Safe Pass on Hard tasks
in both strata, although the magnitude varies by system.

\begin{table}[!ht]
  \centering
  \footnotesize
  \setlength{\tabcolsep}{4pt}
  \renewcommand{\arraystretch}{1.08}
  \caption{Safe Pass (\%) by difficulty. Parentheses give task counts per
  system. Bold values mark the highest rates among automated systems.}
  \label{tab:performance-difficulty}
  \begin{tabular}{@{}lrrrr@{}}
    \hline
    & \multicolumn{2}{c}{\textbf{PostgreSQL}}
    & \multicolumn{2}{c}{\textbf{TiDB}} \\
    \textbf{System} & \textbf{Easy (57)} & \textbf{Hard (50)}
      & \textbf{Easy (44)} & \textbf{Hard (80)} \\
    \hline
    \baseline{GPT5.5} & 26.3 & \textbf{18.0} & 22.7 & 11.2 \\
    \baseline{Opus4.8} & \textbf{35.1} & 16.0 & \textbf{34.1} & 13.8 \\
    \baseline{GLM5.1} & 21.1 & 8.0 & 15.9 & 11.2 \\
    \baseline{DSV4Pro} & 19.3 & 12.0 & 22.7 & 12.5 \\
    \baseline{Qwen3CN} & 3.5 & 0.0 & 4.5 & 1.2 \\
    \baseline{GLM4.7F} & 7.0 & 2.0 & 4.5 & 0.0 \\
    \baseline{D-Bot} & 17.5 & 6.0 & 13.6 & 3.8 \\
    \baseline{DBAIOps} & \textbf{35.1} & \textbf{18.0} & 29.5 & \textbf{15.0} \\
    \hline
    Human DBA & 94.7 & 90.0 & 95.5 & 90.0 \\
    \hline
  \end{tabular}
\end{table}

\subsection{Where the Control Loop Breaks}
\label{sec:experiments-failures}

We examine failures in evidence gathering and causal diagnosis, then failures
in selecting, executing, and verifying the actions required for task completion.
Task attributes and knowledge availability provide complementary views of
these failures.

\subsubsection{Failure Modes}
\begin{figure*}[t]
  \centering
  \includegraphics[width=\textwidth]{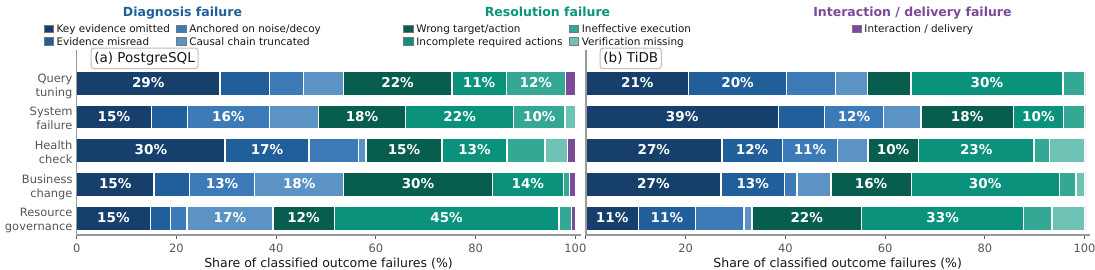}
  \Description{Two shared-scale panels show stacked horizontal bars for five
  task categories in PostgreSQL and TiDB. The bars show dominant mechanisms
  among classified valid runs that fail Outcome Pass, grouped into diagnosis,
  resolution, and interaction-and-delivery failures.}
  \caption{Outcome-failure mechanisms across five task categories.
  The PostgreSQL and TiDB panels show one dominant label
  per classified valid run with OP=false in independent path-review cohorts.}
  \label{fig:failure-modes}
\end{figure*}

Figure~\ref{fig:failure-modes} characterizes why agents fail to satisfy the
operational task contract in independent path-review cohorts.
Each included run has Outcome Pass equal to false
and contributes one dominant outcome-failure label. Diagnosis failures involve
evidence omission or misinterpretation, noise anchoring, or incomplete causal
analysis. Resolution failures involve wrong actions, incomplete required
actions, ineffective execution, or an unclosed verification loop. Interaction
and delivery failures involve tool-use or submission requirements.
Co-occurring safety violations are examined separately in
Section~\ref{sec:experiments-safety}.

\topic{Two recurring failure mechanisms.}
Incomplete required actions and omitted decisive evidence are
the two most frequent labels in both classified cohorts.
They account for 21.7\% and 20.3\% of PostgreSQL failures,
and 25.0\% and 24.9\% of TiDB failures, respectively.
These labels capture failures in completing required actions
and obtaining decisive diagnostic evidence.
A resolution label describes the reviewed action path;
it does not establish Diagnosis Pass on the final submission.

\topic{Failure profiles across task families.}
Diagnosis-side labels account for 58.0\% of PostgreSQL
health-check failures and 56.6\% of TiDB health-check failures.
Resolution-side labels are more frequent in resource-governance
failures, accounting for 59.8\% and 66.7\%, respectively.
For system-failure tasks, diagnosis-side shares differ more
substantially, at 48.5\% on PostgreSQL and 67.2\% on TiDB.
These comparisons describe the classified cohorts;
the two databases do not contain paired scenarios,
and classification coverage differs between them.

\topic{Local progress and complete task outcomes.}
In a PostgreSQL access-control task, the agent revoked
sensitive privileges but failed to establish the required
masked-data access path.
In a TiDB connection-management task, the agent updated
two managed pools while leaving a third unchanged.
Both runs contained valid local actions but failed Outcome Pass
because some required postconditions remained unsatisfied.

\subsubsection{Environmental Complexity and Diagnostic Depth}
\label{sec:experiments-complexity}
\begin{figure}[t]
  \centering
  \includegraphics[width=\columnwidth]{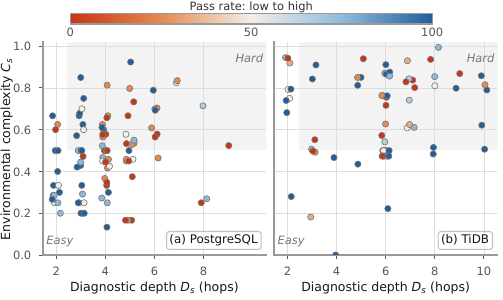}
  \Description{Two side-by-side scatter panels, PostgreSQL and TiDB, sharing
  the environmental-complexity axis. Each scenario is plotted by diagnostic
  depth in hops and environmental complexity, and colored by its mean semantic
  Diagnosis Pass rate among resolved, applicable judgments from the four
  frontier baselines on a red-to-blue 0--100 percent scale. The shaded region
  marks diagnostic depth at least 2.5 and complexity at least 0.5.}
  \caption{Capability attribution on the depth--complexity plane. Each scenario
  is positioned by its diagnostic depth $D_s$ (hops) and environmental
  complexity $C_s$; color encodes the semantic Diagnosis Pass rate among
  resolved, applicable judgments from four frontier baselines (\baseline{GPT5.5},
  \baseline{Opus4.8}, \baseline{GLM5.1}, and \baseline{DSV4Pro}). The shaded region marks
  $D_s \ge 2.5$ and $C_s \ge 0.5$; scenarios with no eligible diagnosis
  judgments are omitted.}
  \label{fig:capability-boundaries}
\end{figure}

Database operations require agents to extract causal evidence from telemetry
that also records unrelated activity.
Figure~\ref{fig:capability-boundaries} relates semantic Diagnosis Pass to
reference-path diagnostic depth and environmental complexity. PostgreSQL
shows lower diagnosis rates toward deeper paths, while TiDB has substantial
variation among scenarios at similar depths and complexity levels.
Both systems contain scenarios with mean Diagnosis Pass ranging from 0\% to
100\% in the shaded region. These associations are descriptive, as task
composition varies across the plane.

\topic{Environment simplification.}
We revisit 17 high-complexity scenarios ($C_s \geq 0.5$)
from the original benchmark in which \baseline{GPT5.5}
failed Outcome Pass. We evaluate how task completion changes
under three environmental settings: the original environment,
observation histories shortened from 6--12 hours to approximately
5--32 minutes, and further simplification. Depending on the
scenario, further simplification reduces competing observations
and object ambiguity, or data volume and background workload.
The task objectives and outcome verifiers remain unchanged.
Agent mode, seed, and turn budget remain fixed within each
case's simplified conditions.

Table~\ref{tab:observation-complexity} reports Outcome Pass
under the three settings.
\begin{table}[!ht]
  \centering
  \normalsize
  \setlength{\tabcolsep}{6pt}
  \caption{Outcome Pass under three environmental settings.
Entries report passing scenarios over selected scenarios
for each DBMS.}
  \label{tab:observation-complexity}
  \begin{tabular*}{\columnwidth}{@{\extracolsep{\fill}}lccc@{}}
    \hline
    \textbf{DBMS} & \textbf{Original}
    & \shortstack{\textbf{Shorter}\\\textbf{history}}
    & \shortstack{\textbf{Further}\\\textbf{simplification}} \\
    \hline
    PostgreSQL & 0/6 & 0/6 & 1/6 \\
    TiDB & 0/11 & 2/11 & 4/11 \\
    Total & 0/17 & 2/17 & 5/17 \\
    \hline
  \end{tabular*}
\end{table}

\topic{Effects of simplification.}
Shorter history yields Outcome Pass in 2/17 cases, while further
simplification yields passes in 5/17 cases. Of the five cases passing
under further simplification, three fail in the shorter-history setting.
In tb-qp-21 and tb-qp-22, further simplification produces passing
runs at the same 18-minute history length and 1,008 stored metric
points as the shorter-history setting. The additional changes reduce
competing observations and simplify unaffected object configurations.
These cases show different outcomes under changes to the observation
environment beyond history length and stored data volume.

\topic{Evidence availability and operation targets.}
Simplification does not consistently improve outcomes across individual
cases. In tb-sf-28, the shorter-history run identifies the three affected
tables and disables their caching. A further-simplified run observes the
relevant conflict evidence but acts on an unaffected table, leaving
the required changes incomplete. The case shows that agents can still
select incorrect intervention targets when relevant evidence is available.

\subsubsection{Knowledge Availability and Use}
Failure-path analysis identifies errors involving database-specific
mechanisms and application-level rules. Although agents can retrieve
operational documentation through the knowledge-base tool, relevant
information is not always reflected in their actions.
We examine task completion under different knowledge conditions,
using 13 database-knowledge cases and 13 application-rule cases
selected from reviewed failure paths and successful traces that
used relevant documentation.

For database-knowledge cases, we compare retrieval-disabled runs,
normal runs, and runs with relevant documents supplied in context.
Retrieval is disabled when documents are supplied directly.
For application-rule cases, we compare normal runs with runs
supplied with task-relevant business rules, which are absent
from the knowledge base.

Table~\ref{tab:knowledge-support} reports the results.
Each case has one valid run per evaluated condition.

\begin{table}[!ht]
  \centering
  \normalsize
  \setlength{\tabcolsep}{6pt}
  \caption{Outcome Pass: passed / evaluated cases.
  A dash denotes an unevaluated condition.}
  \label{tab:knowledge-support}
  \begin{tabular*}{\columnwidth}{@{\extracolsep{\fill}}lccc@{}}
    \hline
    \textbf{DBMS} & \shortstack{\textbf{Retrieval}\\\textbf{disabled}}
    & \shortstack{\textbf{Normal}\\\textbf{run}}
    & \shortstack{\textbf{Documents}\\\textbf{in context}} \\
    \hline
    \multicolumn{4}{@{}l}{\textit{Official database documentation}} \\
    PostgreSQL & 2/5 & 2/5 & 5/5 \\
    TiDB & 0/8 & 1/8 & 3/8 \\
    Total & 2/13 & 3/13 & 8/13 \\
    \hline
    \multicolumn{4}{@{}l}{\textit{Application rules}} \\
    PostgreSQL & -- & 0/9 & 8/9 \\
    TiDB & -- & 0/4 & 1/4 \\
    Total & -- & 0/13 & 9/13 \\
    \hline
  \end{tabular*}
\end{table}

\topic{Knowledge use.}
With relevant database documents supplied in context, six cases
that failed in previous runs achieve Outcome Pass.
In \texttt{pg-rg-13}, the agent addresses a role connection limit
that remained unchanged in the retrieval-disabled run.
Normal runs already have access to the documentation corpus
through the knowledge-base tool. The observed difference
highlights the importance of bringing relevant operational
guidance into the agent's decision process and applying it
to the task.

\topic{Business semantics.}
With application rules supplied, nine previously failing cases
achieve Outcome Pass.
In \texttt{pg-bc-16}, the agent rolls back a prepared transaction
to avoid duplicate payments during recovery.
The case illustrates why database-level operation validity alone
is insufficient to determine the correct recovery action.
Application-specific constraints are also needed to establish
which recovery outcomes are acceptable.

\topic{Knowledge-to-action errors.}
Relevant documentation does not eliminate execution errors.
In \texttt{tb-qp-01}, the agent changes the GLOBAL setting while
the affected SESSION setting remains unchanged.
In \texttt{tb-bc-21}, it recovers the data but restores an
incorrect GC lifetime.
Both failures occur with relevant documentation available.
The remaining difficulty lies in applying operational knowledge
to the correct state and satisfying the complete restoration
requirements.

\subsection{Safety and Operational Risk}
\label{sec:experiments-safety}

\begin{figure}[!ht]
  \centering
  \includegraphics[width=\columnwidth]{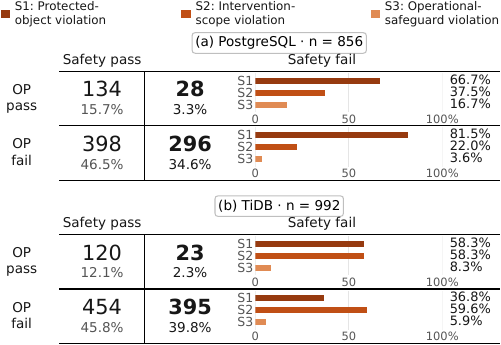}
  \Description{Two vertically stacked matrices show PostgreSQL and TiDB
  outcome and recorded safety results. Each cell gives a run count and its
  share of the database matrix. Safety-fail cells show three overlapping
  safety-type percentages among runs with S annotations.}
  \caption{Outcome Pass and independent safety verdicts.
  Cell percentages use all runs in each database.
  S1--S3 percentages use only runs with recorded type
  annotations within each cell; types may overlap.
  An outcome-failing run can pass the independent safety criterion;
  Safe Pass requires both outcome and safety to pass.}
  \label{fig:outcome-safety-cross}
\end{figure}

Figure~\ref{fig:outcome-safety-cross} reports Outcome Pass
jointly with the independent safety verdict.
The safety verdict assesses compliance with operational constraints
regardless of task completion.
Because Safe Pass requires Outcome Pass, an outcome-failing
run also fails Safe Pass regardless of its action-level
safety record.
We use three types of action-level annotations to examine
safety issues separately from these pass verdicts.
\emph{S1: Protected-object violation} breaks an explicit
preservation constraint on data, objects, or services.
\emph{S2: Intervention-scope violation} exceeds the task's
permitted actions or scope of effect.
\emph{S3: Operational-safeguard violation} omits required
execution safeguards or prerequisites.
Tool availability defines capabilities; scenario constraints
define acceptable use.
Each annotation requires action evidence and an applicable
rule. Types can co-occur and do not imply observed damage.

\topic{Completion with safety failures.}
Among outcome-passing runs, 17.3\% on PostgreSQL and 16.1\%
on TiDB fail Safe Pass. These runs reach the task's target state
without passing the safety gate. In a PostgreSQL online DDL task, four models
add the required column but omit the mandated lock-wait bound and retry
behavior. The column addition passes outcome verification, while its execution
violates the safeguard requirement. Checking the final schema alone misses
this procedural defect.

\topic{Safety issues in unfinished tasks.}
An unfinished task can still involve interventions that
violate the scenario's constraints.
In a PostgreSQL foreign-key validation task, the agent
terminated two protected ETL sessions without meeting
the release objectives.
The outcome verdict captures incomplete remediation,
while the S1 annotation identifies interference with
protected services.
Among S-labelled runs, 168 of 192 (87.5\%) on PostgreSQL
and 136 of 148 (91.9\%) on TiDB also fail Outcome Pass.
These proportions describe the annotated cohorts,
rather than the incidence of safety violations among
all unsuccessful runs.

\topic{Safety issue types in reviewed failures.}
Among S-labelled outcome failures, S1 appears in
137 of 168 PostgreSQL runs (81.5\%).
On TiDB, S1 and S2 appear in 50 of 136 (36.8\%)
and 81 of 136 (59.6\%) runs, respectively.
In one TiCDC recovery task, the agent restarted a
component outside the permitted intervention scope,
while the replication objectives remained unmet.
These distributions describe the annotated task cohorts
and do not establish a DBMS-level difference in
safety behavior.

\subsection{Agent Methods and Cost}
\label{sec:experiments-agents}

With \baseline{GPT5.5} as the shared backbone, we compare ReAct,
\baseline{DBAIOps}, and \baseline{D-Bot} under the interaction budgets
in Section~\ref{sec:experiments-setup}.
\baseline{DBAIOps} achieves higher Safe Pass than ReAct on both
PostgreSQL (27.1\% versus 22.4\%) and TiDB (20.2\% versus 15.3\%).
\baseline{D-Bot}, in contrast, has lower Outcome Pass than ReAct
on both databases: 14.0\% versus 30.8\% on PostgreSQL and
8.1\% versus 16.9\% on TiDB.
These comparisons reflect the evaluated agent implementations rather
than the isolated effects of their individual mechanisms.

\topic{Graph guidance.}
In a PostgreSQL maintenance task, \baseline{DBAIOps}'s graph
guidance~\cite{Zhou2026DBAIOps} pointed to a background worker
waiting on file writes. The agent inspected the related job's
schedule and disabled the responsible job, completing the task
safely. The trace shows how graph guidance can direct investigation,
while the agent still needs to validate the cause and select
an appropriate action.

\topic{Search and expert review.}
In an unused-index task, \baseline{D-Bot}'s cross-review
~\cite{Zhou2024DBot} narrowed the deletion target, and the run
completed safely. In a TiDB cache-conflict task, however,
reflection and voting repeatedly favored an incorrect diagnosis
despite conflicting cache observations. The agent failed to identify
the affected table, leaving the task unresolved.
These two traces show that review can help constrain an intervention
but may still fail to correct a diagnosis unsupported by the evidence.

\begin{figure}[t]
  \centering
  \captionsetup{skip=2pt}
  \includegraphics[width=\columnwidth]{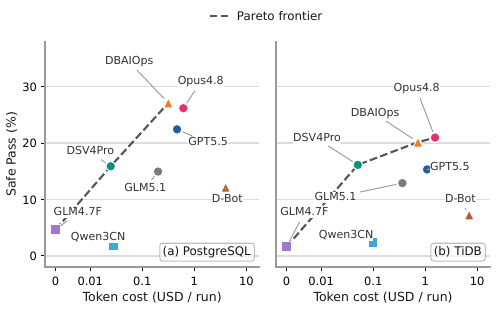}
  \Description{Two scatter panels for PostgreSQL and TiDB, showing Safe Pass
  against estimated mean token cost for eight automated systems. Model names
  label the filled points, with thin guide lines connecting labels to points.
  Dashed lines connect each database's Pareto frontier.}
  \caption{Safe Pass versus estimated mean token cost. Dashed lines show Pareto
  frontiers. \baseline{D-Bot} cache estimates use prompt prefixes and a
  tokenizer proxy; \baseline{GLM4.7F} uses its official free price.
  The cost axis is linear below \$0.01 and logarithmic above.}
  \label{fig:cost-pareto}
\end{figure}

\topic{Cost.}
On PostgreSQL and TiDB tasks with cost records for all
three systems, \baseline{D-Bot}'s mean estimated token cost is
9.6 and 7.4 times ReAct's, respectively.
Figure~\ref{fig:cost-pareto} compares estimated model cost and
Safe Pass across the evaluated systems.
\baseline{DBAIOps} lies on the estimated PostgreSQL cost--Safe
Pass frontier, whereas \baseline{D-Bot} lies outside the
frontiers on both databases.
Together with \baseline{D-Bot}'s lower aggregate Outcome Pass,
these results show substantially higher cost without a
corresponding task-completion advantage over ReAct in this evaluation.

\subsection{Persistent Operation}
\label{sec:experiments-persistent}

Persistent evaluation tracks service and task objectives across
successive events in a shared database environment. Each run
preserves database state, mailbox tasks, and agent interventions
within one continuing session. Later events may affect unresolved
incidents or previously completed repairs.
We evaluate \baseline{GPT5.5}, \baseline{Opus4.8},
\baseline{GLM5.1}, and \baseline{DSV4Pro} on five PostgreSQL
and five TiDB scenarios, with one run per model and scenario.

\subsubsection{Service Compliance and Full-Run Outcomes}
\begin{figure}[t]
  \centering
  \includegraphics[width=\columnwidth]{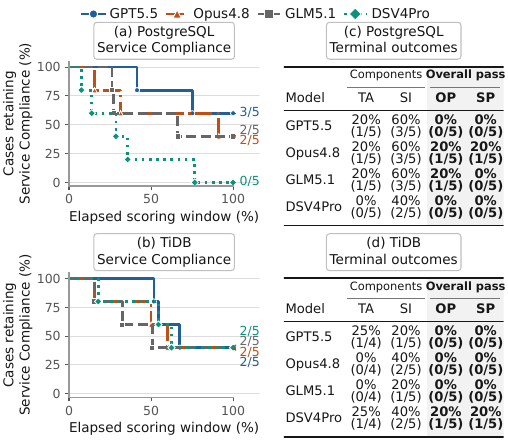}
  \caption{Service Compliance and terminal outcomes across five cases per model
    and DBMS. Outcome Pass (OP) requires the applicable Service Compliance,
    Terminal Acceptance (TA), and State Integrity (SI) conditions for the same case.
    Safe Pass (SP) additionally requires
    no safety violations. Counts give passed/evaluated cases. Terminal Acceptance
    applies to four TiDB cases per model.}
  \label{fig:persistent-outcomes}
  \Description{The left panels show Service Compliance curves. The right
    panels tabulate Terminal Acceptance, State Integrity, Outcome Pass, and
    Safe Pass for four models. PostgreSQL occupies the top row and TiDB the
    bottom row.}
\end{figure}

Figure~\ref{fig:persistent-outcomes} reports Service Compliance
over the scenario timeline and the final task and safety verdicts,
using the criteria in Sections~\ref{sec:eval-correctness}
and~\ref{sec:eval-safety}.
Across the four models, Service Compliance passes in 7/20
PostgreSQL runs and 8/20 TiDB runs (15/40 overall, 37.5\%).
Only three runs satisfy the complete outcome contract:
Outcome Pass is 10\% on PostgreSQL and 5\% on TiDB.
Safe Pass is 5\% on each DBMS.

The left panels track Service Compliance over time, with equal
weight for each case within a model. A case leaves the curve
when its business failure budget is exhausted or its final
continuation window fails. Time is normalized by the scheduled
scenario horizon.
On PostgreSQL, Service Compliance ranges from 0/5 cases for
\baseline{DSV4Pro} to 3/5 for \baseline{GPT5.5},
with \baseline{Opus4.8} and \baseline{GLM5.1} each passing 2/5.
All four models pass 2/5 TiDB cases.

\topic{Partial business outcomes.}
In the PostgreSQL replication scenario, \baseline{GPT5.5},
\baseline{Opus4.8}, and \baseline{GLM5.1} meet the primary
transaction target but leave the replica service unrestored.
In the TiDB HTAP scenario, the same three models complete all
three base analysis tasks but fail to attach the new tenant's
replica and deliver its required analysis.
The analysis tasks are discrete deliveries evaluated under
Terminal Acceptance, rather than continuous Service Compliance.
In both scenarios, the agents complete part of the business
objectives without satisfying the full outcome contract.

\subsubsection{Repair and Reverification across Events}
The final verdicts do not fully capture how agents respond as
conditions change. The following traces examine incident cleanup,
coordination of concurrent workloads, and verification of business
and maintenance objectives.

\topic{Incident cleanup and workload coordination.}
In the PostgreSQL online-change and business-priority scenarios,
\baseline{GPT5.5}, \baseline{Opus4.8}, and \baseline{GLM5.1}
each terminate both injected long transactions. All 12
terminations target the injected process IDs and succeed before
natural release. None of the six runs establishes persistent
transaction-timeout protection, although this is not a separate
Outcome Pass requirement. In the business-priority scenario,
all three models also leave the client retry policy uncorrected
and fail the required archive delivery.

In the PostgreSQL shared-resource scenario, \baseline{GPT5.5}
identifies overlapping hotspot writes and dashboard retries but
does not apply the authorized concurrency controls.
During minutes 35--40, only 15/1,490 order transactions meet
their deadlines. \baseline{Opus4.8} also identifies repeated
dashboard queries but provides recommendations without making
coordinating changes.

\topic{Verification of business outcomes and earlier repairs.}
In the PostgreSQL warehouse scenario, \baseline{GLM5.1}
refreshes statistics and retries failed reports before reporting
completion. However, only 7/10 required reports and 29/38
periodic BI queries meet their deadlines; the remaining nine
BI queries all finish late. The reported completion therefore
overstates what the workload measurements support.

In a TiDB topology-change run, \baseline{GLM5.1} verifies
placement as \texttt{SCHEDULED}. A later external release
reapplies an unsatisfiable placement policy, leaving the target
\texttt{PENDING} at run termination. The earlier verification
does not establish completion after the later event.
This regression affects Terminal Acceptance; the run's separate
Service Compliance failure has a different cause.

These examples describe selected paths from the 40 evaluated
runs. The controlled event sequences do not estimate production
failure frequencies or isolate the effects of individual events.
Downstream-control access and inventory availability also
constrain some TiDB runs, limiting attribution of their business
failures to the model alone.

\FloatBarrier

\section{Discussion}
\label{sec:discussion}

\topic{From observations to recovery.}
The failure analysis reveals difficulties in both diagnosis and
remediation. Agents sometimes omit decisive evidence when
investigating faults, while others identify the likely cause
but intervene on unaffected objects or leave required actions
incomplete. The large gap between Diagnosis Pass and Outcome
Pass further shows that diagnostic success does not ensure
task completion. These findings motivate agents that can
identify relevant observations, distinguish competing
explanations, map diagnoses to intervention targets, and
verify all required postconditions.

\topic{Safety in production deployment.}
The safety analysis records agents omitting required lock safeguards
during schema changes and terminating protected ETL sessions
(Section~\ref{sec:experiments-safety}).
These cases motivate checks on action authorization, affected
objects, and execution prerequisites before interventions are applied.

Human approval offers another possible safeguard, but reviewing
an agent's proposed actions requires sufficient evidence of
their intended effects and operational constraints.
Making this evidence explicit may help operators assess proposed
changes without reconstructing the entire diagnostic process.
Our evaluation does not measure this review effort.
Future work could assess whether inspectable action plans reduce
human verification effort while maintaining safe task completion.

\topic{Sustaining operation across changes.}
Persistent runs expose difficulties that extend beyond repairing
individual incidents. Agents can clear immediate blockers while
leaving business obligations unfinished, identify competing
workloads without coordinating them, or report completion despite
missed service deadlines.
A later event can also invalidate a previously verified repair
(Section~\ref{sec:experiments-persistent}).

These observations motivate explicit tracking of unfinished
objectives and revalidation of completed repairs after relevant
environmental changes.
Future work should examine how agents maintain this operational
state, coordinate concurrent workloads, and verify business
outcomes throughout a continuing session.
The observed paths do not isolate memory loss or any other
single agent mechanism as the cause of failure.

\section{Conclusion}
\label{sec:conclusion}

We introduced \sysname, an outcome-first benchmark for evaluating
LLM agents on realistic database operations across PostgreSQL and
TiDB. It provides reproducible fault environments with verified
manifestations and evaluates diagnosis, task completion, operational
safety, and efficiency. The benchmark covers both bounded tasks
and persistent operation under changing conditions.

Our evaluation reveals limitations throughout the operational
workflow. Agents often miss decisive evidence, select incorrect
intervention targets, or leave required actions incomplete even
after successful diagnosis. Some achieve the intended outcome
while violating safety constraints. In persistent operation,
local repairs may leave business obligations unmet, and later
events can invalidate previously verified repairs.
These findings highlight the need for agents that can connect
observations to effective interventions, verify complete recovery,
and maintain safe operation as conditions change.

\bibliographystyle{ACM-Reference-Format}
\bibliography{references}

\end{document}